\documentclass[twocolumn,conference,10pt]{IEEEtran}
\usepackage{epsfig,amsfonts,subfigure}
\usepackage[nolist]{acronym}
\usepackage{graphicx,cite,amssymb,amsmath}
\usepackage{color}
\usepackage{tikz}
\usetikzlibrary{shapes,arrows}

\IEEEoverridecommandlockouts

\begin{document}
\begin{acronym}
\acro{LT}{Luby-transform} \acro{CER}{codeword error rate}
\acro{ML}{maximum-likelihood}\acro{MPE}{multi-protocol
encapsulation}\acro{FEC}{forward error
correction}\acro{ADT}{application data table}\acro{RS}{Reed
Solomon}\acro{ARQ}{automatic retransmission query}\acro{LRFC}{linear
random fountain code}\acro{MDS}{maximum distance
separable}\acro{SPC}{single-parity-check}
\end{acronym}

\tikzstyle{block} = [draw, fill=white!20, rectangle, minimum
    height=3em, minimum width=7em, text width=6em, text centered]
\tikzstyle{point} = [draw, fill=black!20, circle]
\tikzstyle{input}=[coordinate] \tikzstyle{output} = [coordinate]
\tikzstyle{left_of_block} = [coordinate] \tikzstyle{left_of_LRFC}
=[coordinate] \tikzstyle{r_input} =[coordinate]

\tikzstyle{pinstyle} = [pin edge={to-,thin,black}]

\title{On the Concatenation of Non-Binary Random Linear Fountain Codes with Maximum Distance Separable Codes}
 \author{\authorblockN{Francisco Lazaro Blasco}
\authorblockA{Institute of Communications\\
and Navigation\\
DLR (German Aerospace Center)\\
Wessling, Germany 82234\\
Email: Francisco.LazaroBlasco@dlr.de}
 \and
\authorblockN{Gianluigi Liva}
\authorblockA{Institute of Communications\\
and Navigation\\
DLR (German Aerospace Center)\\
Wessling, Germany 82234\\
Email: Gianluigi.Liva@dlr.de} } \maketitle
\date{\today}
\thispagestyle{empty} \setcounter{page}{0}


\begin{abstract}
\textcolor{black}{The performance of a novel fountain coding scheme
based on maximum distance separable (MDS) codes constructed over
Galois fields of order $q\geq2$ is investigated. Upper and lower
bounds on the decoding failure probability under maximum likelihood
decoding are developed. \textcolor{black}{Differently from Raptor
codes (which are based on a serial concatenation
\textcolor{black}{of} a high-rate outer block code, and an inner
Luby-transform code),} \textcolor{black}{the proposed coding scheme
can be seen as a parallel concatenation} of an outer MDS code and an
inner random linear fountain code, both operating on the same Galois
field.  A performance assessment is performed on the gain provided
by MDS based fountain coding over linear random fountain coding in
terms of decoding failure probability vs. overhead. It is shown how,
for example, the concatenation of a $(15,10)$ Reed-Solomon code and
a linear random fountain code over $\mathbb {F}_{16}$ brings to a
decoding failure probability $4$ orders of magnitude lower than the
linear random fountain code for the same overhead in a channel with
a packet loss probability of $\epsilon=5\cdot10^{-2}$. Moreover, it
is illustrated how the performance of the concatenated fountain code
approaches that of an idealized fountain code for higher-order
Galois fields and moderate packet loss probabilities.
\textcolor{black}{The scheme introduced is of special interest for
the distribution of data using small block sizes.}}
\end{abstract}

{\pagestyle{plain} \pagenumbering{arabic}}


\section{Introduction}\label{sec:Intro}
\textcolor{black}{ Fountain codes were introduced in
\cite{byers02:fountain} as an efficient alternative to \ac{ARQ}
protocols in multicast/broadcast transmission systems. Consider the
case where a sender (or source) needs to deliver a file to a set of
$N_u$ users. Consider furthermore the case where users are affected
by packet losses. In this scenario, the usage of an \ac{ARQ}
protocol can result in large inefficiencies, since users may loose
different packets, and hence a large number of retransmissions would
crowd the downlink channel. Among the efficient (coded) alternatives
to \ac{ARQ} protocols
\cite{Metzer84:retransmission,GLCK_Patent_2008,Medard08:ARQ,Sorour09:ARQ},
we shall focus on fountain codes only. When a fountain code is used,
the source file is split in a set of $k$ source packets.
\textcolor{black}{The sender, or fountain encoder, computes linear
combinations of the $k$ source packets and broadcasts them through
the communication medium. After receiving $k$ fountain coded
packets, receivers can try to recover the source packets. If they
fail to recover the source packets they will try again to decode
when they receive additional packets.}  The efficiency of a fountain
code deals with the amount of packets (source+redundancy) that a
user needs to collect for recovering the source file. An
\emph{idealized} fountain code would allow the file recovery with a
probability of success $P_s=1$ from any set of $k$ received packets.
Real fountain decoders need in general to receive a larger amount of
packets, $m=k+\delta$, for achieving a certain success probability.
Commonly, $\delta$ is referred to as \emph{overhead} of the fountain
code, and is used to measure its efficiency. \textcolor{black}{More
generally a \emph{universal} fountain code is a code which can
recover the $k$ original source symbols out of $k+\delta$ symbols
for any erasure channel and $\delta$ small. The first class of
universal fountain codes are \ac{LT} codes \cite{luby01:LT}. One
sub-class of \ac{LT} codes are random \ac{LT} codes or \acp{LRFC}
\cite{shokrollahi06:raptor}}. When a binary \ac{LRFC} is used
\cite{MacKay05:fountain,Liva10:fountain} the success probability can
be accurately modeled as $P_s=1-2^{-\delta}$ for $\delta>2$ (it can
be proved that $P_s$ is actually always lower bounded by
$1-2^{-\delta}$, \cite{Liva10:fountain}). In \cite{Liva10:fountain}
it was shown that this expression is still accurate for fountain
codes based on sparse matrices (e.g., Raptor codes
\cite{shokrollahi06:raptor}). \textcolor{black}{Moreover,} in
\cite{Liva10:fountain}, the performance achievable by performing
linear combinations of packets on Galois fields of order greater
than $2$ was analyzed. For a \ac{LRFC} performing the linear
combinations over $\mathbb {F}_q$, the decoding failure probability
$P_e=1-P_s$ is bounded by \cite{Liva10:fountain}
\begin{equation}\label{eq:tightbounds}
q^{-\delta-1}\leq P_e(\delta,q) < \frac{1}{q-1}q^{-\delta}
\end{equation}
where both bounds are tight for increasing $q$.
\textcolor{black}{Furthermore,} in \cite{Liva10:fountain} it was
\textcolor{black}{also} shown that non-binary Raptor codes can in
fact tightly approach the bounds \eqref{eq:tightbounds} down to
moderate error rates.}

 \textcolor{black}{The result is remarkable, considering that for a Raptor code the
encoding and decoding costs (defined as the number of arithmetic
operations divided by the number of source symbols, $k$) are
\textcolor{black}{$\mathcal O( \hspace{1pt} \log(1/\alpha) )$ and
$\mathcal O(k \hspace{1pt} \log(1/\alpha))$} respectively, being
$k(1+\alpha)$ the number of output symbols needed to recover the
source symbols with a high probability. For a \ac{LRFC} the encoding
cost is $\mathcal O(k)$ and the decoding cost is $\mathcal O(k^2)$,
and thus it does not scale favorably with the input block size.
However, if the block size is kept small, the decoding cost is still
affordable.}

 \textcolor{black}{The motivation of this paper is the analysis of a further
improvement of the approach proposed in \cite{Liva10:fountain} for
designing fountain codes with good performance for short block
sizes. As in \cite{Liva10:fountain}, in order to achieve the
objective non-binary fountain codes are considered. Moreover,
\ac{MDS} codes are introduced in parallel concatenation with the
fountain encoder to enhance the performance of the scheme. By doing
that, the first $n$ output symbols of the encoder are the $n$ output
symbols of the \ac{MDS} code.\footnote{Note that for Raptor codes
the output of the precode is further encoded by a \ac{LT} Code.
Hence the first $n$ output symbols of the fountain encoder are not
the output of the precode.}\footnote{We will assume a \ac{MDS}
linear block code constructed on the same field ($\mathbb {F}_q$) of
the fountain code.} \newline In this paper, we illustrate how the
performance of \acp{LRFC} \textcolor{black}{in terms of probability
of decoding failure} can be further improved by such a
concatenation.  An analytical expression for the decoding failure
probability vs. overhead will be derived under the assumption of
\ac{ML} decoding. We show how, when the packet loss rates are
moderate-low, the probability of failure can be reduced
\textcolor{black} {by} several orders of magnitude, approaching the
performance of idealized fountain codes. The simulated performance
of schemes based on \ac{RS} codes are compared with the proposed
expressions, confirming the accuracy of the proposed approach. The
analysis is developed for the case \textcolor{black}{of} \acp{LRFC}.
We conjecture that similar gains shall be expected also in the case
where (non-binary) \ac{LT} codes are employed in the concatenation.}

The paper is organized as follows. In Section
\ref{sec:concatenation} the proposed concatenated scheme is
introduced. In Section \ref{sec:bounds} the performance analysis is
provided, while numerical results are presented in Section
\ref{sec:results}. Conclusions follow in Section \ref{sec:conc}.

\section{Concatenation of Block Codes with Random Linear Fountain
Codes}\label{sec:concatenation}

\begin {figure}[h]
\begin{center}
{\small
\begin{tikzpicture}[auto, node distance=2cm, label distance=6mm, >=latex']
    \node [input, name=input] {};
    \node [r_input, right of=input, node distance = 1.5cm](r_input) {};
    \node [left_of_block, above of =r_input, node distance = 1.5 cm] (left_of_block) {};
    \node [left_of_LRFC, below of =r_input, node distance = 1.5 cm] (left_of_LRFC) {};
    \node [block, right of=left_of_block,  node distance = 2 cm] (blockcode) {Block Code  $(n,k)$};
    \node [block, right of=left_of_LRFC, , node distance = 2 cm] (LRFC) {LRFC};
    \node [point, right of=blockcode, node distance = 2 cm] (blockcode_point){};
    \node [point, right of=LRFC, node distance = 2 cm] (LRFC_point) {};
    \node [point, right of=input, node distance = 7cm] (out_point) {};
    \node [output, right of=out_point, node distance = 1.5 cm] (output) {};
    \draw [-]  (input) -- node [label=above:{$u_1,u_2...u_k$}] { } (r_input);
    \draw [-]  (r_input) -- node  { } (left_of_block);
    \draw [-]  (r_input) -- node { } (left_of_LRFC);
    \draw [->] (left_of_block) -- node {} (blockcode);
    \draw [->] (left_of_LRFC) -- node {} (LRFC);
    \draw [-] (blockcode) -- node
    [label=above:{$c_1,c_2...c_n$}] {} (blockcode_point);
    \draw [-] (LRFC) -- node
    [ label=above:{$c_{n+1},c_{n+2}...$}] {} (LRFC_point);
    \draw [->] (out_point) -- node
    [label=above:{$c_1,c_2...c_n,c_{n+1}...$}] {} (output);
    \draw [-] (blockcode_point) -- node{} (out_point);
\end{tikzpicture}}
\caption{Fountain coding scheme seen as a parallel concatenation of
a $(n,k)$ linear block code and a linear random fountain
code.}\label{fig:par}
\end{center}
\end {figure}
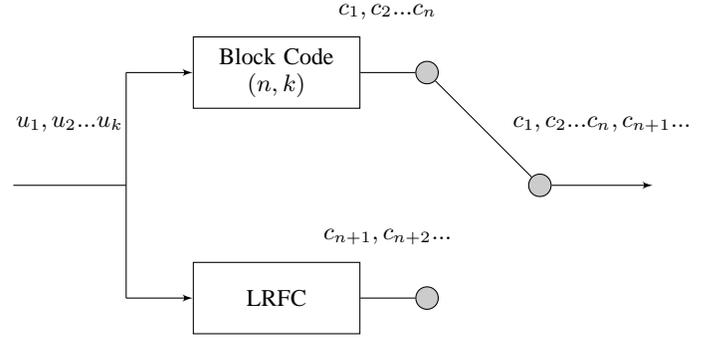

\textcolor{black}{Without loosing in generality, we define the
source block $\mathbf{u}=(u_1, u_2, \ldots, u_k)$ as a sequence of
symbols belonging to a Galois field of order $q$, i.e.
$\mathbf{u}\in \mathbb {F}_q^k$. In the proposed approach, the
source block is first encoded via a $(n,k)$ systematic linear block
code $\mathcal{C}'$ over $\mathbb {F}_q$ with generator matrix
$\mathbf{G}'=(\mathbf{I}|\mathbf{P}')$, where $\mathbf{I}$ is the
$k\times k$ identity matrix and $\mathbf{P}'$ is a $k\times (n-k)$
matrix with elements in $\mathbb {F}_q$. The encoded block is hence
given by
$\mathbf{c}'=\mathbf{u}\mathbf{G}'=(c'_1,c'_2,\ldots,c'_n)$, where
$c'_1=u_1, c'_2=u_2,  \ldots, c'_k=u_k$ and the remaining $n-k$
symbols of $\mathbf{c}'$ are the redundancy symbols given by
$(c'_{k+1},c'_{k+2},\ldots,c'_n)=\mathbf{u}\mathbf{P}'$. Additional
redundancy symbols can be obtained by computing random linear
combinations of the $k$ source symbols as
\begin{displaymath}
c_i=c_{i-n}''=\textcolor{black} {\sum_{j=1}^{k}}g_{j,i}u_j, \qquad
i=n+1,\ldots, l
\end{displaymath}
where the coefficients $g_{j,i}$ are picked from $\mathbb {F}_q$
with a uniform probability ($1/q$). The encoded sequence is hence
given by $\mathbf{c}=(\mathbf{c}'|\mathbf{c}'')$. The overall
generator matrix has the form
\begin{equation}
\mathbf{G}=
\underbrace{\left(\begin{array}{cccc}
  g_{1,1} & g_{1,2} & \ldots & g_{1,n} \\
  g_{2,1} & g_{2,2} & \ldots & g_{2,n} \\
  \vdots  & \vdots  & \ddots & \vdots  \\
  g_{k,1} & g_{k,2} & \ldots & g_{k,n}
\end{array}\right|}_{\mathbf{G}'}  \underbrace{\left|\begin{array}{cccc}
  g_{1,n+1} & g_{1,n+2} & \ldots  & g_{1,l} \\
  g_{2,n+1} & g_{2,n+2} & \ldots  & g_{2,l} \\
  \vdots    & \vdots    & \ddots  & \vdots  \\
  g_{k,n+1} & g_{k,n+2} & \ldots  & g_{k,l}
\end{array}\right)}_{\mathbf{G}''}
\end{equation}
where $\mathbf{G}''$ is the generator matrix of the \ac{LRFC}. (Note
that, being the \ac{LRFC} rate-less, the number $l$ of columns of
$\mathbf{G}$ can in principle grow indefinitely.) The encoder can be
seen hence as a parallel concatenation of the linear block code
$\mathcal C '$ and of a \ac{LRFC} (Fig. \ref{fig:par}).}

\section{Performance Analysis} \label{sec:bounds} Based on the bounds
derived in \cite{Liva10:fountain}, tight upper and lower bounds for
the decoding failure probability of the fountain coding scheme can
be derived in case of uncorrelated erasures. The decoding failure
probability ($P_F=\textrm{Pr}\{F\}$, where $F$ denotes the decoding
failure event) is defined as the probability that the source block
$\mathbf{u}$ cannot be recovered out of a set of received symbols.
In this paper we will focus on the case where the linear block code
used in concatenation with the \ac{LRFC} is maximum distance
separable (MDS). When binary codes will be used, we will assume
$(k+1,k)$ \ac{SPC} codes.\footnote{Repetition codes are not
considered here, since they would lead to a trivial fountain scheme
where the source block is given by $1$ symbol only.} When operating
on higher order Galois fields, we will consider (shortened) \ac{RS}
codes.

\textcolor{black}{The encoded sequence is given by
$\mathbf{c}=\mathbf{u}\mathbf{G}=(c_1,c_2,\ldots,c_l)$, where the
first $n$ symbols $(c_1,c_2,\ldots,c_n)$ represent a codeword of
$\mathcal C'$, and the remaining $l-n$ are produced by the
\ac{LRFC}. At the receiver side, a subset of $m$ symbols is
received. We denote by $J=\{j_1, j_2, \ldots, j_{m}\}$ the set of
the indexes on the symbols of $\mathbf{c}$ that have been received.
The received vector \textcolor{black}{$\mathbf{y}$} is hence given
by
\[
\mathbf{y}=(y_1, y_2, \ldots,
y_{m})=(c_{j_1},c_{j_2},\ldots,c_{j_{m}})
\]
and it can be related to the source block $\mathbf{u}$ as
$\mathbf{y}=\mathbf{u}\tilde{\mathbf{G}}$. Here,
$\tilde{\mathbf{G}}$ denotes the $k\times m$ matrix made by the
columns of $\mathbf{G}$ with indexes in $J$, i.e.
\[
\tilde{\mathbf{G}}= \left(\begin{array}{cccc}
  g_{1,j_1} & g_{1,j_2} & \ldots & g_{1,j_{m}} \\
  g_{2,j_1} & g_{2,j_2} & \ldots & g_{2,j_{m}} \\
  \vdots  & \vdots  & \ddots & \vdots  \\
  g_{k,j_1} & g_{k,j_2} & \ldots & g_{k,j_{m}}
\end{array}\right).
\]
The recovery of $\mathbf{u}$ reduces to solving the system  of
$m=k+\delta$ linear equations in $k$ unknowns
\begin{equation}
\tilde{\mathbf{G}}^T\mathbf{u}^T=\mathbf{y}^T,\label{eq:solve}
\end{equation}
 e.g., via Gaussian
elimination. The solution is possible if and only if $\textrm{rank}
(\tilde{\mathbf{G}})=k$. }

\textcolor{black}{Assuming $\mathcal C'$ being \ac{MDS}, the system
is solvable with probability $1$ if, among the $m$ received symbols,
at least $k$ have indexes in $\{1, 2, \ldots, n\}$, i.e. if at least
$m'\geq k$ symbols produced by the linear block encoder have been
received.}

\textcolor{black}{Let's consider the less trivial case where $m'<k$
among the $m$ received symbols have indexes in $\{1, 2, \ldots,
n\}$. We can partition $\tilde{\mathbf{G}}^T$ as
\begin{equation}
\tilde{\mathbf{G}}^T=\left(\begin{array}{c} \tilde{\mathbf{G}}'^T\\
\tilde{\mathbf{G}}''^T \end{array}\right)=\left(\begin{array}{cccc}
  g_{1,j_1} & g_{2,j_1} & \ldots & g_{k,j_1} \\
  g_{1,j_2} & g_{2,j_2} & \ldots & g_{k,j_2} \\
  \vdots  & \vdots  & \ddots & \vdots  \\
  g_{1,j_{m'}} & g_{2,j_{m'}} & \ldots & g_{k,j_{m'}}\\ \hline
  g_{1,j_{m'+1}} & g_{2,j_{m'+1}} & \ldots & g_{k,j_{m'+1}} \\
  g_{1,j_{m'+2}} & g_{2,j_{m'+2}} & \ldots & g_{k,j_{m'+2}} \\
  \vdots  & \vdots  & \ddots & \vdots  \\
  g_{1,j_{m}} & g_{2,j_{m}} & \ldots & g_{k,j_{m}}
\end{array}\right). \label{eq:G_partition}
\end{equation}
The \ac{MDS} property of $\mathcal{C}'$ assures that $\textrm{rank}
(\tilde{\mathbf{G}}')=m'$, i.e. the first $m'$ rows of
$\tilde{\mathbf{G}}^T$ are linearly independent. Note that the
$m''\times k$ matrix $\tilde{\mathbf{G}}''^T $ (with $m''=m-m'$) is
a random matrix whose entries are picked with uniform probability in
$\mathbb F _q$. It follows that the system defined by
\eqref{eq:G_partition} can be put (via column permutations over
$\tilde{\mathbf{G}}^T$ and row permutations/combinations over
$\tilde{\mathbf{G}}'^T$) in the form
\begin{equation}
\hat{\mathbf{G}}^T=\left(\begin{array}{ccc} \mathbf{I} & \vline &
\mathbf{A} \\\hline
\mathbf{0} & \vline & \mathbf{B}\\
\end{array}\right), \label{eq:G_partition_manipulation}
\end{equation}
where $\mathbf I$ is the $m' \times m'$ identity matrix,
$\mathbf{0}$ is a $m'' \times m'$ all-$0$ matrix, and  $\mathbf{A}$,
$\mathbf{B}$ have respective sizes $m' \times (k-m')$ and $m''
\times (k-m')$. Note that the lower part of $\hat{\mathbf{G}}^T$
given by $\left(\mathbf{0} \,\, \mathbf{B}\right)$ is obtained by
adding to each row of $\tilde{\mathbf{G}}''^T$ a linear combination
of rows from $\tilde{\mathbf{G}}'^T$, in a way that the $m'$
leftmost columns of $\tilde{\mathbf{G}}''^T$ are zeroed-out. It
follows that the statistical properties of $\tilde{\mathbf{G}}''^T$
are inherited by the $m'' \times (k-m')$ sub-matrix $\mathbf{B}$,
whose entries are hence picked with uniform probability in $\mathbb
F_q$. The system is solvable if and only if $\mathbf{B}$ is full
rank, i.e. if and only if $\textrm{rank}(\mathbf{B})=k-m'$.}

\textcolor{black}{Suppose now that the encoded symbols $\mathbf{c}$
are sent to a receiver over an erasure channel which erasure
probability of $\epsilon$. The probability that at least $k$ symbols
out of the $n$ symbols produced by the linear block code encoder are
received is given by
\begin{equation}
Q^*(\epsilon)=\sum_{i=k}^n {n \choose i} (1-\epsilon)^i
\epsilon^{n-i}.
\end{equation}
Hence, with a probability $P^*(\epsilon)=1-Q^*(\epsilon)$ the
receiver would need to collect symbols encoded by the \ac{LRFC}
encoder to recover the source block. Assuming that the user collects
$m=k+\delta$ symbols, out of which only $m'<k$ have been produced by
the linear block encoder, the conditional decoding failure
probability can be expressed as
\begin{equation}
\textrm{Pr}(F|m',m'<k,\delta)=\textrm{Pr}(\textrm{rank}(\mathbf{B})<k-m').\label{eq:cond_F_prob_1}
\end{equation}
Note that $\mathbf{B}$ is a $m'' \times (k-m') = (k+\delta-m')
\times (k-m')$ random matrix, i.e. a random matrix with $\delta$
equations in excess w.r.t. the number of unknowns. We can thus
replace \eqref{eq:tightbounds} in \eqref{eq:cond_F_prob_1}, getting
the bounds
\begin{equation}
q^{-\delta-1}\leq\textrm{Pr}(F|m',m'<k,\delta)\frac{1}{q-1}q^{-\delta}.\label{eq:cond_F_prob_2_bounds}
\end{equation}
We remark that, thanks to the independency of the bounds in
\eqref{eq:tightbounds} from the size of the random matrix (i.e., the
bounds depend only on the overhead), we can remove the conditioning
on $m'$ from \eqref{eq:cond_F_prob_2_bounds}, leaving
\[
q^{-\delta-1}\leq\textrm{Pr}(F|m'<k,\delta)<\frac{1}{q-1}q^{-\delta}.
\]
The final failure probability can be written as
\begin{equation}
\begin{array}{cc}
P_F(\delta,\epsilon)=& \textrm{Pr}(F|m'<k,\delta)\textrm{Pr}(m'<k)+\\
&+\textrm{Pr}(F|m'\geq k,\delta)\textrm{Pr}(m'\geq k),
\label{eq:general_bound}
\end{array}
\end{equation}
where $\textrm{Pr}(F|m'\geq k,\delta)=0$ and
$\textrm{Pr}(m'<k)=P^*(\epsilon)$. It results that
\begin{equation}
P^*(\epsilon) q^{-\delta-1}\leq
P_F(\delta,\epsilon)<P^*(\epsilon)\frac{1}{q-1}q^{-\delta}.\label{eq:final_bounds}
\end{equation}
From an inspection of \eqref{eq:tightbounds} and
\eqref{eq:final_bounds}, one can note how the bounds on the failure
probability of the concatenated scheme are scaled down by a factor
$P^*(\epsilon)$, where $P^*(\epsilon)=\sum_{i=0}^{k-1} {n \choose i}
(1-\epsilon)^i \epsilon^{n-i}$ is a monotonically increasing
function of $\epsilon$. It follows that, when the channel conditions
are \emph{bad} (i.e., large $\epsilon$) $P^*(\epsilon)\rightarrow
1$, and the bounds in \eqref{eq:final_bounds} tend to coincide with
the bounds in \eqref{eq:tightbounds}. When the channel conditions
are \emph{good} (i.e., small $\epsilon$), most of the time $m'\geq
k$ symbols produced by the linear block encoder are received,
leading to a decoding success (recall the assumption of \ac{MDS}
code). In these conditions, $P^*(\epsilon)\ll 1$, and according to
the bounds in  \eqref{eq:final_bounds} the failure probability may
scale down even of  several orders of magnitude.}

\begin{figure}[h]
\begin{center}
\includegraphics[width=0.92\columnwidth,draft=false]{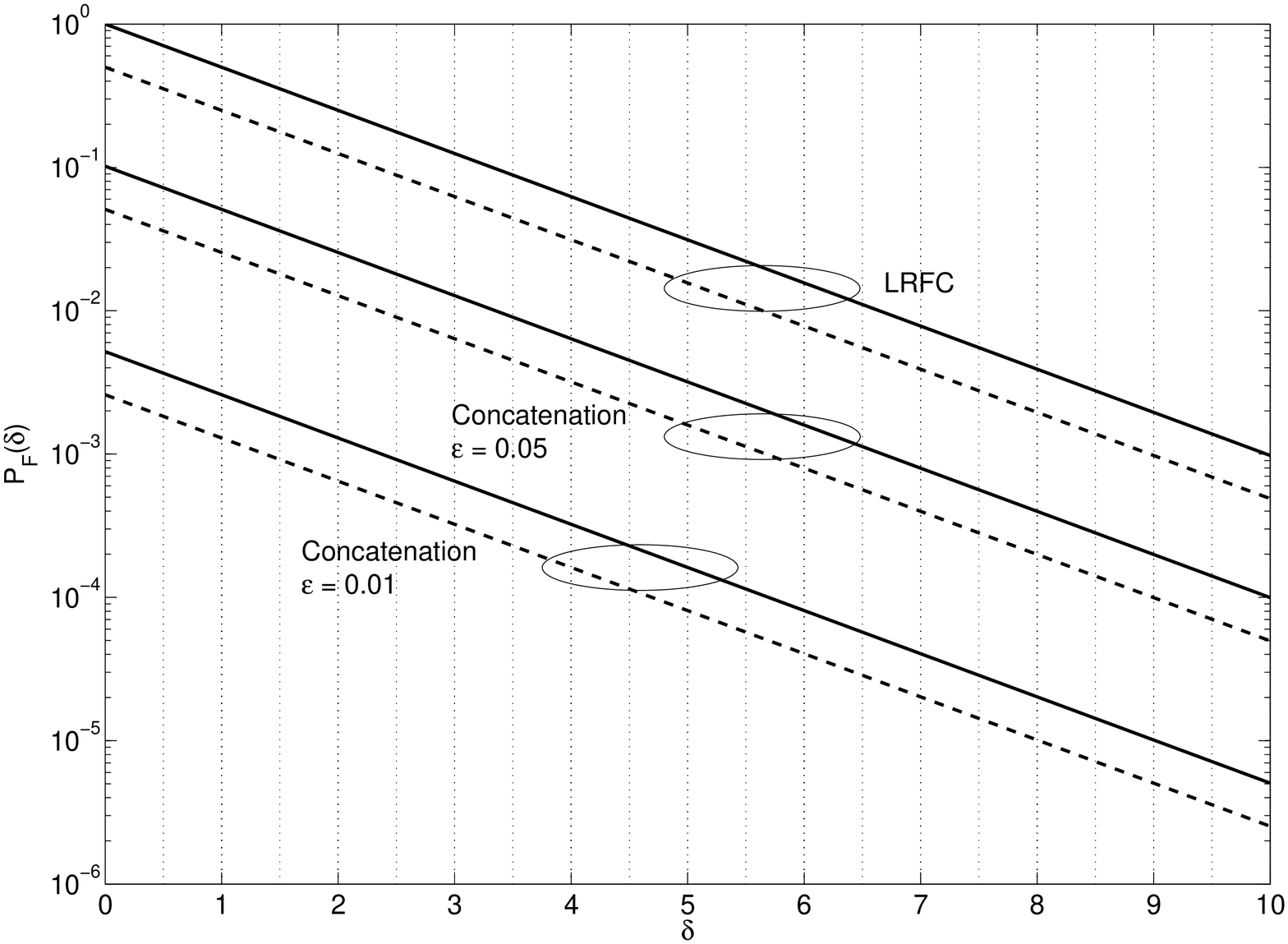}
\centering \caption{$P_F(\delta,\epsilon)$ vs. overhead  for a concatenated code built using a $(11,10)$ \ac{SPC} code over $\mathbb {F}_{2}$ for  different values of $\epsilon$.
Upper bounds are represented by solid lines and lower bounds are represented by dashed lines.} \label{GF_2}
\end{center}
\end{figure}

\begin{figure}[h]
\begin{center}
\includegraphics[width=0.92\columnwidth,draft=false]{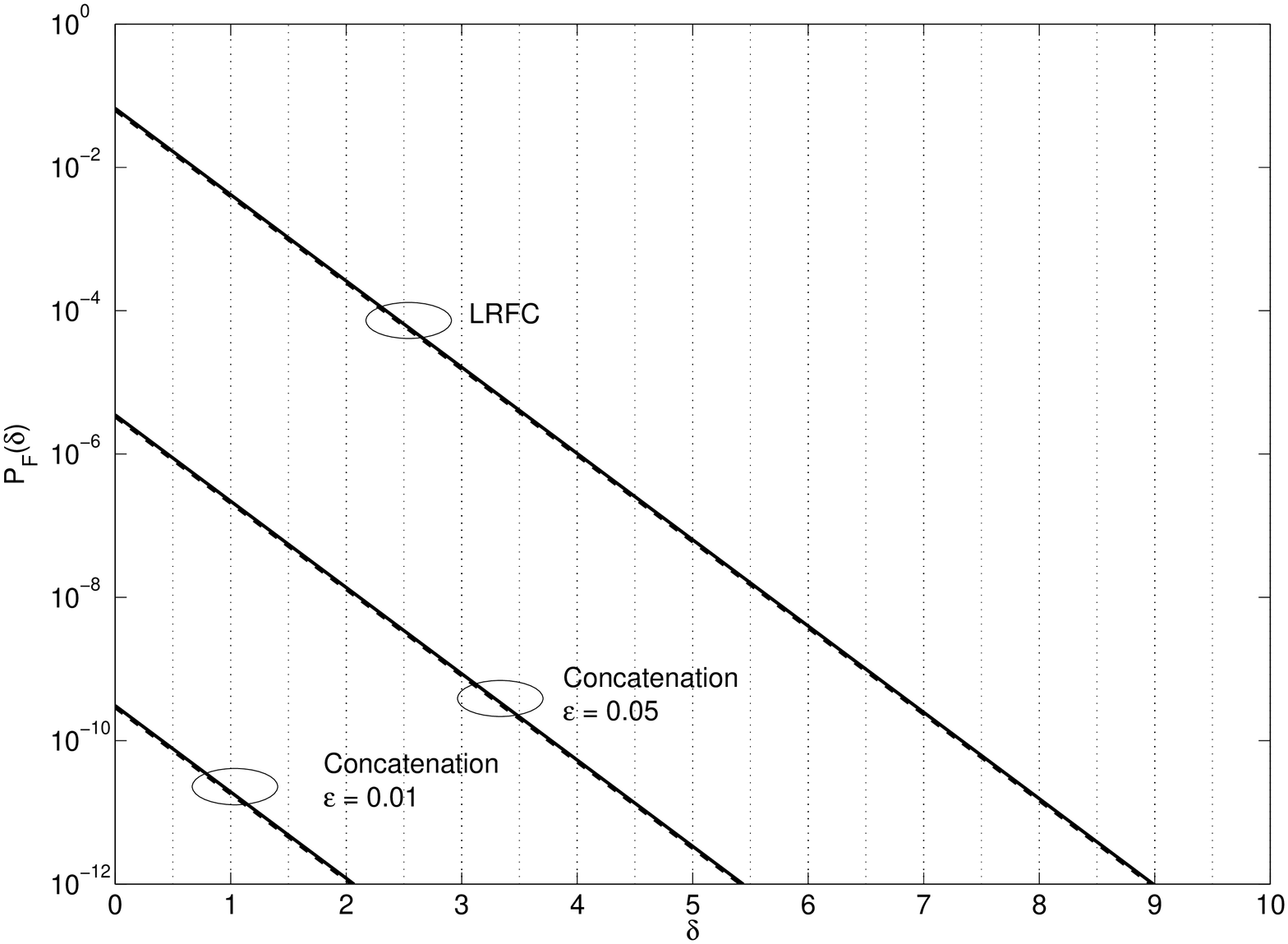}
\centering \caption{$P_F(\delta,\epsilon)$ vs.  overhead  for a concatenated code built using a $(15,10)$ \ac{RS} over $\mathbb {F}_{16}$ for  different values of $\epsilon$.
Upper bounds are represented by solid lines and lower bounds are represented by dashed lines.} \label{GF_16}
\end{center}
\end{figure}

Fig. \ref{GF_2} shows the probability of decoding failure as a
function of the number of overhead symbols for a concatenated code
built using a $(11,10)$ \ac{SPC} code in $\mathbb {F}_2$. It can be
observed how, for lower erasure probabilities, the performance gain
in terms of probability of decoding failure increases. For
$\epsilon=0.01$ the decoding failure probability is more than $2$
orders of magnitude lower. Fig. \ref{GF_16} shows the probability of
decoding failure vs. the number of overhead symbols for the
concatenation of a $(15,10)$ \ac{RS} and a \ac{LRFC} over $\mathbb
{F}_{16}$. The performance of the concatenated code is compared with
that of the \ac{LRFC} built on the same field for different erasure
probabilities. In this case the decrease in terms of probability of
decoding failure is bigger than in for the previously presented code
in $\mathbb {F}_2$. For a channel with an erasure probability
$\epsilon=0.05$, the probability of decoding failure of the
concatenated scheme is $4$ orders of magnitude lower than for the
\ac{LRFC}.

\textcolor{black}{The analysis provided in this section is also
valid if the \ac{LRFC} is substituted by a \ac{LT} or Raptor code.
In order to calculate the performance of such a concatenated code
one has to substitute in \eqref{eq:general_bound} the term
$\textrm{Pr}(F|m'<k,\delta)$ by the probability of decoding failure
of the \ac{LT} or Raptor code. Again the failure probability of the
concatenated scheme is scaled down by a factor $P^*(\epsilon)$,
where $P^*(\epsilon) \leq 1$.}

\section{Numerical Results}\label{sec:results}
\begin{figure}[h]
\begin{center}
\includegraphics[width=0.92\columnwidth,draft=false]{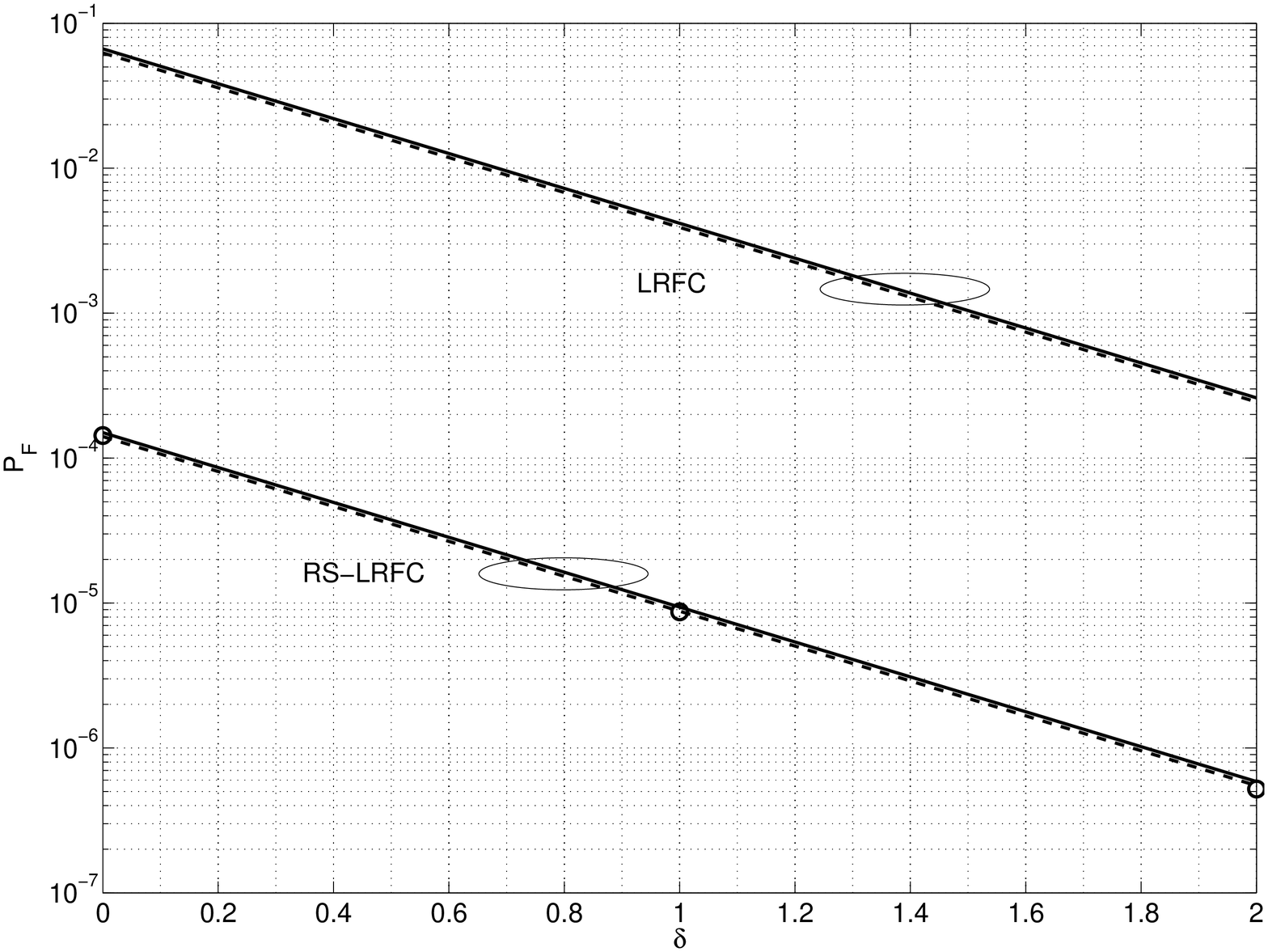}
\centering \caption{$P_F(\delta,\epsilon)$ vs. overhead  for a
the concatenation of a $(15,10)$ \ac{RS} and \ac{LRFC} over $\mathbb {F}_{16}$ and $\epsilon=0.1$. Upper bounds are represented by solid lines and lower bounds are represented by dashed lines. The points marked with '$\circ$' denote actual simulations.} \label{GF_16_sim}
\end{center}
\end{figure}

\begin{figure}[h]
\begin{center}
\includegraphics[width=0.92\columnwidth,draft=false]{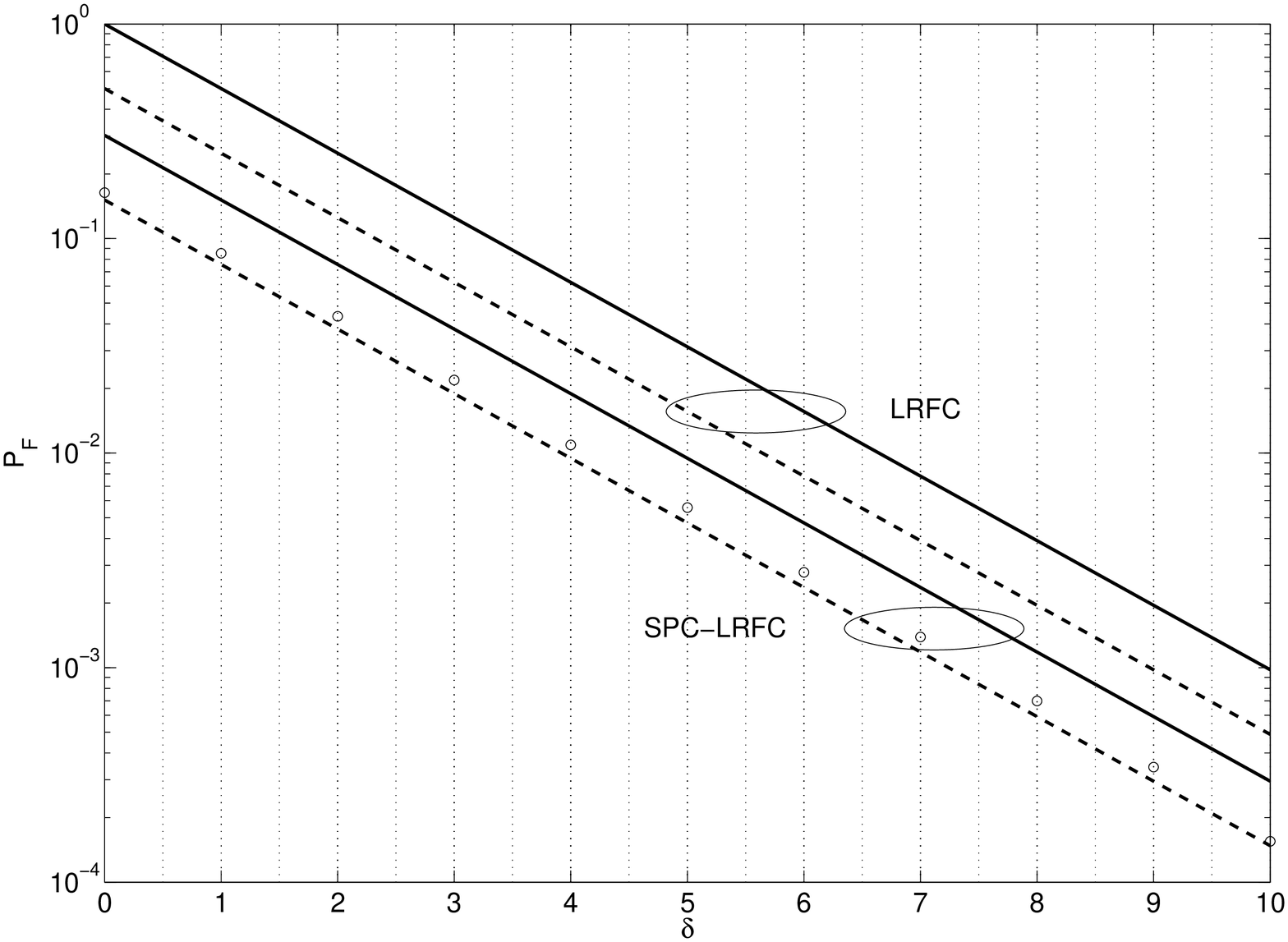}
\centering \caption{$P_F(\delta,\epsilon)$ vs. overhead symbols for a
the concatenation of a $(11,10) $\ac{SPC} code and a \ac{LRFC} over $\mathbb {F}_{2}$ and $\epsilon=0.1$. Upper bounds are represented by solid lines and lower bounds are represented by dashed lines. The points marked with '$\circ$' denote actual simulations.} \label{GF_2_sim}
\end{center}
\end{figure}

Fig. \ref{GF_16_sim} shows the results of simulations together with
the bounds calculated using (\ref{eq:final_bounds}). In this case a
\textcolor{black}{$(15,10)$} \ac{RS} was concatenated with a
\ac{LRFC} in $\mathbb {F}_{16}$, and a channel with an erasure
probability $\epsilon=0.1$ was used. It can be seen how the
simulation results match the analytical results down to a
probability of decoding failure of $10^{-7}$. Fig. \ref{GF_2_sim}
shows the simulation results for a concatenated code using a
$(11,10)$ parity check code in $\mathbb {F}_{2}$, and a channel with
an erasure probability $\epsilon=0.1$. It can be seen how the
simulation results match the analytical results again. However, in
$\mathbb {F}_{2}$ the bounds are less tight than in higher order
Galois fields.

An assessment the performance of the concatenated scheme in a system
with a high number of users has been performed, assuming a system in
which a transmitter sends a source block to a set of $N$ receivers.
We considered the erasure channels from the transmitter to the
receivers to be independent, with an identical erasure probability
$\epsilon$. Furthermore, we assumed that the receivers send an
acknowledgement to the transmitter when they have successfully
decoded the block. Ideal (error-free) feedback channels have been
considered. When all receivers have sent an acknowledgement, the
transmitter stops encoding redundant symbols for the source block.

\begin{figure}[h]
\begin{center}
\includegraphics[width=0.92\columnwidth,draft=false]{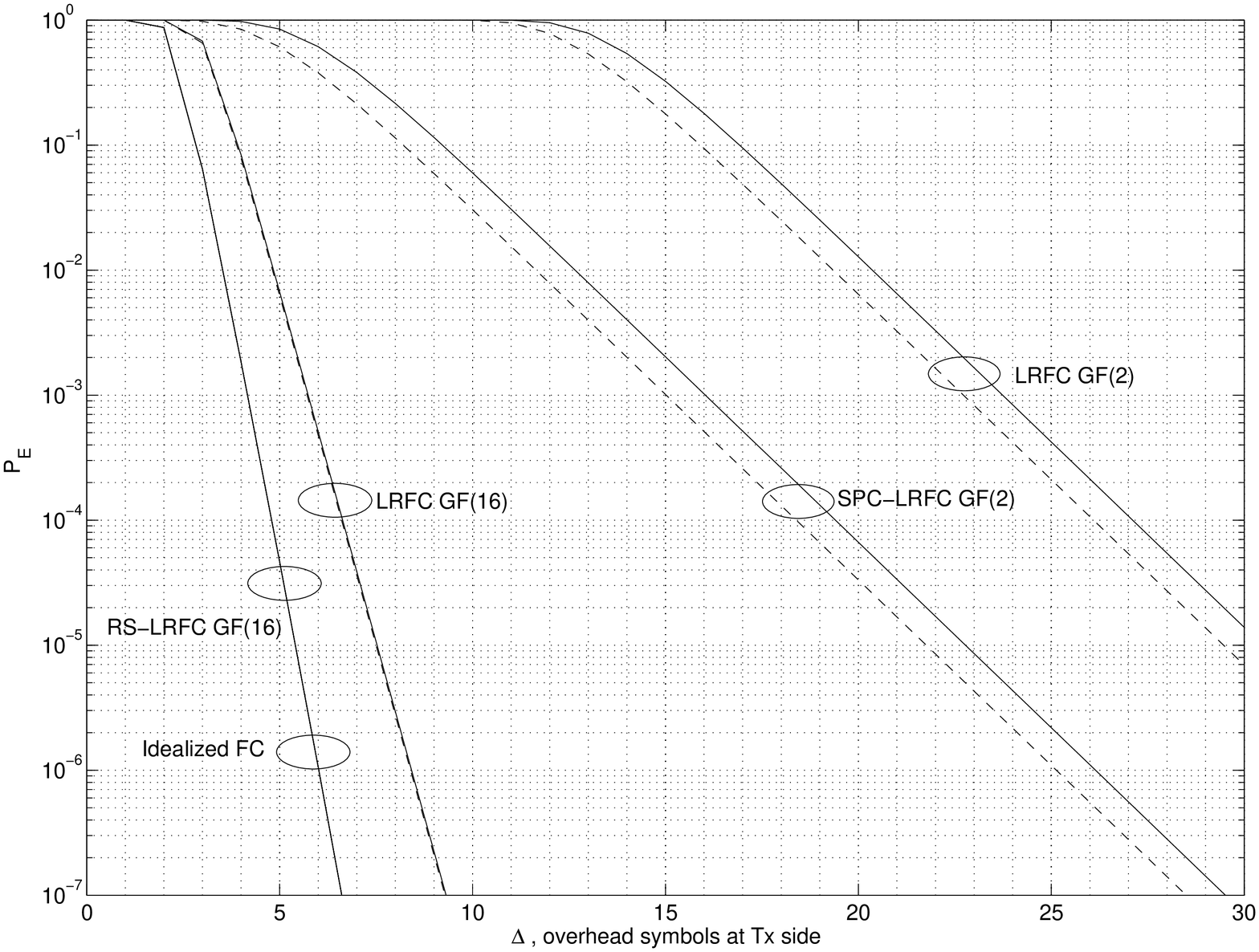}
\centering \caption{ $P_E$ vs.overhead at the transmitter in a system with $N=10000$ users and $\epsilon=0.01$ . Results are shown for different fountain codes:
\ac{LRFC} in $\mathbb {F}_{2}$, \ac{LRFC} in $\mathbb {F}_{16}$, concatenation of a (11,10) \ac{SPC} code with a \ac{LRFC} code in $\mathbb {F}_{2}$,
and a concatenation of a \textcolor{black}{$(15,10)$} \ac{RS} code and a \ac{LRFC} code over $\mathbb {F}_{16}$.}\label{sim_sender_side}
\end{center}
\end{figure}

If $k+\Delta$ (where $\Delta$ denotes the transmitter overhead)
symbols have been transmitted, the probability that a specific
receiver gathers exactly $m$ symbols is:
\begin{equation}
\ P_{R}\{k+\Delta,m\} = \binom{k+\Delta}{m}(1-\epsilon)^{m}\epsilon^{k+\Delta-m}
\label{system_prob m}
\end{equation}
The probability of decoding failure at the receiver given that the
transmitter has sent $k+\Delta$ symbols is hence
\begin{align*}
\ P_{e} =& \sum_{m=0}^{k-1}\ P_{R}\{k+\Delta,m\}+\\
& + \sum_{m=k}^{k+\Delta}\ P_{R}\{k+\Delta,m\} P_{F}\{\delta=m-k,\epsilon\}.
\end{align*}
The probability that at least one user has not decoded successfully
is thus
\begin{equation}
\ P_E(N,\Delta,\epsilon) = 1-(1-P_{e})^{N}
\label{system_failure_one user}
\end{equation}
Using the bounds in (\ref{eq:final_bounds}) $P_E(N,\Delta,\epsilon)$
can also be bounded. In the following we provide an example to asses
the performance of the new scheme in comparison with \ac{LRFC} codes
and also with an \emph{idealized} fountain code. We assume a system
with $N=10^{4}$ users and a channel with an erasure probability
$\epsilon=0.01$. The performance of \ac{LRFC} codes over $\mathbb
{F}_{2}$ and $\mathbb {F}_{16}$ is shown as well as that of two
concatenated schemes: a concatenation of a $(11,10)$ \ac{SPC} code
with a \ac{LRFC} code in $\mathbb {F}_{2}$, and a concatenation of a
$(15,10)$ \ac{RS} code and a \ac{LRFC} code over $\mathbb {F}_{16}$.
It can be seen how the concatenated scheme in $\mathbb {F}_{2}$
outperforms the \ac{LRFC} constructed on the same Galois field. For
example, for $P_E = 10^{-4}$ the concatenated scheme in $\mathbb
{F}_{2}$ needs only $\Delta=20$ overhead symbols whereas the
\ac{LRFC} needs $27$ (Fig. \ref{sim_sender_side}). In the case of
the fountain codes operating in $\mathbb {F}_{16}$, the concatenated
code shows a performance very close to that of an \emph{idealized}
fountain code.

\section{Conclusions}\label{sec:conc}
A novel fountain coding scheme has been introduced. The scheme
consists of a parallel concatenation of a \ac{MDS} block code with a
\ac{LRFC} code, both constructed over the same field, $\mathbb
{F}_{q}$. The performance of the concatenated fountain coding scheme
has been analyzed through derivation of tight bounds on the
probability of decoding failure as a function of the overhead. It
has been shown how the concatenated scheme performs as well as
\ac{LRFC} codes in channels characterized by high erasure
probabilities, whereas they provide failure probabilities lower by
several orders of magnitude at moderate/low erasure probabilities.

\section{Acknowledgments}
The authors would like to acknowledge Dr. Enrico Paolini, Dr.
Francesco Rossetto and Giuliano Garrammone for the useful
discussions.

\bibliography{IEEEabrv,studio3}


\end{document}